\documentstyle[12pt,epsf]{article}
\def\beq{\begin{equation}}
\def\eeq{\end{equation}}

\def\beqn{\begin{eqnarray}}
\def\eeqn{\end{eqnarray}}
\def\ra{\rightarrow}
\def\prd#1#2#3{{\it Phys. Rev.} {\bf D#1} #2 (19#3)}
\def\pl#1#2#3{{\it Phys. Lett.} {\bf #1B} #2 (19#3)}
\def\np#1#2#3{{\it Nucl. Phys.} {\bf B#1} #2 (19#3)}
\def\prl#1#2#3{{\it Phys. Rev. Lett.} {\bf #1} #2 (19#3)}
\def\cp{{\cal CP}}
\def\pp{\overline{p}}

\def\tmis{\not{\,T}}
\relax
\begin{document}
\begin{titlepage}
\def\ba{\begin{array}}
\def\ea{\end{array}}
\def\thefootnote{\fnsymbol{footnote}}
\vfill
\hskip 4in {BNL-HET-SD-96-6}

\hskip 4in {ISU-HET-96-2}

\hskip 4in {September 1996}
\vspace{1 in}
\begin{center}
{\large \bf $\cp$ VIOLATION IN $W\gamma$ and $Z\gamma$ PRODUCTION}\\

\vspace{1 in}
{\bf S.~Dawson,$^{(a)}$}\footnote{This manuscript has been authored
under contract number DE-AC02-76CH00016 with the U.S. Department
of Energy.  Accordingly, the
U.S. Government retains a non-exclusive, royalty-free license to
publish or reproduce the published form of this contribution, or
allow others to do so, for U.S. Government purposes.} 
{\bf  Xiao-Gang~He$^{(b)}$} 
{\bf  and G.~Valencia$^{(c)}$}\\
{\it  $^{(a)}$ Physics Department,
               Brookhaven National Laboratory,  Upton, NY 11973}\\
{\it  $^{(b)}$ School of Physics,
               University of Melbourne,
               Parkville, Vic 3052, Australia}\\
{\it  $^{(c)}$ Department of Physics and Astronomy,
               Iowa State University,
               Ames IA 50011}\\
\vspace{1 in}
\end{center}
\begin{abstract}

We study the capability of a $2~{\rm TeV}$ 
$p\overline{p}$ collider with an integrated luminosity of $10~fb^{-1}$ 
to study $\cp$ violation in the processes 
$p \bar{p} \ra W^\pm \gamma$ and $p \bar{p} \ra Z \gamma$. We assume the 
existence of new $\cp$ violating interactions beyond the standard model 
which we describe with an effective Lagrangian. We find that the study of 
$\cp$-odd observables would allow this machine to place bounds on 
$\cp$ violating anomalous couplings similar to the bounds that the same 
machine can place on $\cp$ conserving anomalous couplings. For example 
it could place the bound $|\tilde{\kappa}_\gamma|<0.1$ at the $95\%$ 
confidence level.

\end{abstract}

\end{titlepage}

\clearpage

\section{Introduction}

There has been a recent effort to investigate the physics potential 
of a high-luminosity upgrade of the Fermilab Tevatron collider \cite{tev}. 
One of the directions of this study is the possibility to probe new 
$\cp$ violating interactions in $W$ and $Z$ physics. 
We have previously 
studied the issue of searching for $\cp$ violation in $W$ and 
$W$~jet production \cite{us}. In this paper we extend that study to 
the case of $W\gamma$ and $Z\gamma$ production.

The motivation for this study is, of course, that the origin of $\cp$ 
violation remains unexplained and the question should be pursued 
experimentally wherever possible. The reactions $p\bar{p}\ra W^\pm 
\gamma{\rm ~or~}Z\gamma$ have been studied at the Tevatron and will 
be studied in further detail at an upgraded machine where samples of 
a few thousand events are expected \cite{tev}. The kinematics of these 
reactions, unlike that of single $W^\pm$ or $Z$ production, allows 
triple-product $\cp$ violating correlations to exist. 

It is known that the standard model and minimal extensions (for example 
multi-Higgs models) do not produce sizable $\cp$ violating effects in 
high-energy processes as the ones we discuss in this paper \cite{gaso}. 
We will assume that there are new $\cp$ violating interactions at high 
energy\footnote{As it is necessary in order to explain the baryon 
asymmetry in the universe.}which  manifest themselves at energy scales 
up to a few TeV as $\cp$ violating operators in an effective Lagrangian 
that involves only the fields present in the minimal standard model 
and respects its symmetries 
\cite{barbieri,buchmuller,nacht,burgess,peccei,appel}. 

The $\cp$ violating operators that could lead to a $\cp$-odd observable 
in direct $W\gamma$ and $Z\gamma$ production would also contribute to 
low energy processes where there is no evidence for $\cp$ violation beyond 
that present in the CKM phase of the minimal standard model. However, as 
we argued in Ref.~\cite{us}, there are several reasons why the higher 
energy processes would be more sensitive to certain types of interactions 
than the lower energy counterparts. Among them, that contributions from the 
new operators (associated with physics at some high energy scale $\Lambda$) 
to amplitudes at an energy scale $\mu$ are suppressed by 
powers of $(\mu/\Lambda)^2$. The effects of these operators in direct 
$W^\pm \gamma$ production are thus enhanced by at least a factor of 
$M_W^2/m_\pi^2 \approx 3.5 \times 10^{5}$ over their effects in, say, 
radiative pion decays~\cite{us}. 
It is true that there are stringent indirect bounds 
on some of the new operators from observables such as the electric dipole 
moment of the neutron \cite{marciano}. 
However, these bounds depend on naturalness assumptions and are, therefore, 
complementary to the ones that can be placed in direct $W^\pm \gamma$ 
 and $Z\gamma$ production.

For our study, we will use $\cp$ odd observables already described in the 
literature or very closely related ones \cite{us,barbieri,nacht,brand}. 
We present this analysis as a framework for experimental searches for 
$\cp$ violation. Having a specific parameterization of $\cp$ violating 
interactions it is possible to compare different observables and distributions, 
and in that way distinguish truly $\cp$ violating new physics from potential 
$\cp$ biases of the detectors. 

\section{$p\pp \ra W^\pm \gamma \ra \ell^\pm \nu \gamma$}

This reaction is similar to the process $p\pp \ra W^\pm  {\rm ~jet}$ that we 
considered in Ref.~\cite{us}. In fact, the same operator that we
 considered in 
that case, Eq.~18 of Ref.~\cite{us}, would generate $\cp$-odd asymmetries in 
this process as well. We will focus here on a different operator that generates 
$\cp$-odd asymmetries in this mode through a $\cp$ violating $WW\gamma$ 
coupling. The $\cp$ violating $WW\gamma$ couplings have been traditionally 
parameterized in terms of $\tilde{\kappa}$ and $\tilde{\lambda}$ following 
Ref.~\cite{hagi}. However, within an effective Lagrangian 
description of the symmetry breaking sector of the standard model, the coupling 
$\tilde{\lambda}$ is suppressed with respect to $\tilde{\kappa}$ by a factor 
$\mu^2/\Lambda^2$ for a process with a typical energy scale $\mu$. We denote 
by  $\Lambda$ the scale of the new $\cp$ violating interactions that give 
rise to $\tilde{\kappa}$ and $\tilde{\lambda}$. 
For this reason we concentrate on the coupling $\tilde{\kappa}$ using the 
effective Lagrangian in a notation similar to that of Ref.~\cite{appel}:
\beq
{\cal L} = {v^2 \over \Lambda^2}\biggl(
{1 \over 4} \alpha_{13} gg^\prime \epsilon^{\mu\nu\rho\sigma}B_{\mu\nu}
{\rm Tr}\bigl(TW_{\rho\sigma}\bigr)+{1\over 8}\alpha_{14}g^2
\epsilon^{\mu\nu\rho\sigma}{\rm Tr}\bigl(TW_{\mu\nu}\bigr)
{\rm Tr}\bigl(TW_{\rho\sigma}\bigr)\biggr).
\label{apefl}
\eeq
This differs from Ref.~\cite{appel} in that we have introduced 
a factor $v^2/\Lambda^2$ ($v\approx 246~{\rm GeV}$), for consistency 
with the power counting relevant for this type of new physics \cite{chano}. In 
terms of the conventional notation of Ref.~\cite{hagi} we have \cite{appel}:
\begin{eqnarray}
\tilde{\kappa}_\gamma &=& -{e^2\over s_\theta^2}{v^2 \over \Lambda^2}
\bigl(\alpha_{13}+\alpha_{14}\bigr)
\nonumber \\
\tilde{\kappa}_Z &=& {v^2 \over \Lambda^2}
\biggl({e^2\over c_\theta^2}
\alpha_{13}-{e^2\over s_\theta^2}\alpha_{14}\biggr),
\end{eqnarray}
where $c_\theta=\cos \theta_W, s_\theta=\sin \theta_W$.
The process $p\bar{p}\ra W^\pm \gamma$ is only sensitive to 
$\tilde{\kappa}_\gamma$. 

The new physics that generates these anomalous couplings will in general 
give rise to form factors in the $WW\gamma$ vertex with imaginary (absorptive) 
parts that are typically of the same size as the real parts. These  
absorptive parts combine with the $\cp$ violating couplings to generate 
additional $\cp$ odd observables. To study these observables we will 
assume that there is such an absorptive phase in the $WW\gamma$ vertex 
and parameterize it by $\sin\delta(WW\gamma)$ which we assume to be a number 
of order one.\footnote{See Ref.~\cite{appel} for a calculation of 
such an absorptive phase within a specific model for $\cp$ violation.} 

With all this in mind we proceed to compute the differential cross-section 
for the process 
$\bar{u}(p_u)d(p_d) \rightarrow \ell^-(p_\ell) \nu(p_\nu) \gamma(p_\gamma)$.
 We find a 
$\cp$ violating term linear in $\tilde{\kappa}_\gamma$ that contains a triple 
product. After squaring the matrix element, averaging over initial quark 
color and spin, and summing over the final photon polarization, 
we find for this term:
\begin{eqnarray}
|{\cal M}_{CP}|^2 &=&
{4\over 3} e^2 g^4 |V_{ud}|^2 
\tilde{\kappa}_\gamma { \epsilon(p_u,p_d,p_\ell,p_\gamma)\over (\hat{s}-M^2_W) 
(m_{\ell\nu}^2-M_W^2)^2} \nonumber \\
&\cdot&\biggl[ 
{2 p_\ell\cdot p_u \over 3(p_\gamma-p_u)^2 }
+{p_\ell\cdot p_u + p_\nu\cdot p_d
\over( \hat{s}-M^2_W)}+ 
{p_\nu \cdot p_d \over 3(p_\gamma-p_d)^2 }
\biggr],
\label{wgtripco}
\end{eqnarray}
where ${\hat s}=  (p_u + p_d)^2$ and $m_{\ell\nu}^2=
(p_\ell+p_\nu)^2$.  
We use the 
narrow width approximation
and, therefore, neglect the contributions to $p\pp \ra \ell^- \nu \gamma$ 
from radiative $W$ decays. To justify this approximation 
we restrict our study to the region of phase space where it 
works best, given by the cuts described in Ref.~\cite{baur}. For our numerical 
results we use set B1 of the Morfin-Tung parton distribution functions 
\cite{morftung} 
evaluated at a scale $\mu^2 = M_W^2 + p_{T\gamma}^2$ and the following cuts:
$p_{T\gamma} > 10~{\rm GeV}$, $p_{T\ell} > 20~{\rm GeV}$, $p_{\tmis} > 
20~{\rm GeV}$, 
$|y_\gamma| < 2.4$, $|y_\ell|<3.0$, $\Delta R(\ell\gamma) > 0.7$ and 
$M_T(\ell\gamma,p_{\tmis}) > 90~{\rm GeV}$. 
The first and last two cuts, defined as in 
Ref.~\cite{baur}, suppress the radiative $W$ decays. The other cuts are 
typical Tevatron acceptance cuts \cite{tev}.

The correlation in Eq.~\ref{wgtripco} generates $\cp$-odd and $T$-odd 
observables based on the following triple-product in the lab frame: 
${\vec p}_\ell\cdot ({\vec p}_{\rm \gamma}\times {\vec p}_{\rm beam})$.
We proceed as in Ref.~\cite{us} and construct the $T$-odd observable:
\begin{equation}
A^\pm = \sigma^\pm  [({\vec p}_\gamma\times {\vec p}_{\rm beam} )
\cdot {\vec p}_\ell > 0] -
\sigma^\pm [({\vec p}_\gamma\times {\vec p}_{\rm beam} )
\cdot {\vec p}_\ell < 0]
\label{tripobs}
\end{equation}
where $A^\pm$ refers to the observable for $W^\pm$ events
(or $\ell^\pm \nu$ events). Using Eq.~\ref{tripobs} we 
construct the following $\cp$ odd observables: 
\begin{eqnarray}
{R}_1 & \equiv &
{A^+ - A^- \over {\sigma}^+ + {\sigma}^-} \nonumber \\
{R}_2(y_0) & \equiv &
{ {d A^+\over dy} \mid_{y=y_0}-
{d A^-\over dy}\mid_{y= -y_0} \over
{d {\sigma}^+\over dy}\mid_{y=y_0} +
{d {\sigma}^-\over dy}\mid_{y= -y_0} } 
\label{cpobst},
\end{eqnarray}
where $y$ can be the rapidity of the lepton or the photon (or the $W$). 
Similar observables can be constructed for other distributions \cite{us} 
but we will not consider them in this paper.

When we allow for a non-zero absorptive phase, 
$\sin\delta(WW\gamma)$, there are 
additional $\cp$ violating terms in the differential cross-section. 
They generate a new set of ($T$-even) $\cp$-odd asymmetries and 
following Ref.~\cite{us} we construct the following: 
\begin{eqnarray}
\tilde{R}_1 & \equiv &
{{\sigma}^+ - {\sigma}^- \over {\sigma}^+ + {\sigma}^-} \nonumber \\
\tilde{R}_2(y_0) & \equiv &
{ {d {\sigma}^+\over dy} \mid_{y=y_0}-
{d {\sigma}^-\over dy}\mid_{y= -y_0} \over
{d {\sigma}^+\over dy}\mid_{y=y_0} +
{d {\sigma}^-\over dy}\mid_{y= -y_0} } 
\label{cpobs},
\end{eqnarray}
where ${\sigma}^\pm$ refers to $\sigma(p\pp\ra\ell^\pm\nu \gamma )$.

With the set of cuts discussed above, we find numerically that:
\begin{eqnarray}
\sigma^- &\approx & \bigl(245 -36\tilde{\kappa}_\gamma 
\sin\delta(WW\gamma)\bigr) fb \nonumber \\
\sigma^+ &\approx & \bigl(245 +36\tilde{\kappa}_\gamma 
\sin\delta(WW\gamma)\bigr)  fb \nonumber \\
A^+& = &- A^- \approx  119\tilde{\kappa}_\gamma  fb
\label{wresult}
\end{eqnarray}
and, therefore, $R_1 \approx 0.49 \tilde{\kappa}_\gamma$ and 
$\tilde{R}_1 \approx 0.15 \tilde{\kappa}_\gamma \sin\delta(WW\gamma)$. 
For an integrated luminosity of $10fb^{-1}$ we thus expect some 
2500 $W^- \gamma$ events within the phase space region defined by our cuts 
and this translates into the  $95\%$ confidence level bound:
\begin{equation}
|\tilde{\kappa}_\gamma|  \leq  0.1 
\label{wbound}
\end{equation}
Taking $\Lambda \sim 1~{\rm TeV}$, the bound Eq.~\ref{wbound} translates 
into $|\alpha_{13}+\alpha_{14} | \leq 4$. In a theory where there is 
no suppression of $\cp$ violating interactions with respect to $\cp$ 
conserving ones\footnote{Among other things this would be a theory 
with no custodial $SU(2)$ symmetry.} these couplings could be of order 
one and thus this bound could be significant. However, it is more likely 
that these couplings are much smaller. For example, 
Ref.~\cite{appel} finds in generic technicolor models, $\alpha_{13,14}\sim 
10^{-4}$ ($\tilde{\kappa}_\gamma \sim 8 \times 10^{-6}$). 

It is interesting to compare this bound with the corresponding bound that 
the same machine could place on similar, but $\cp$ conserving, 
anomalous couplings. For example it has been claimed that a $95\%$ 
confidence level bound $|\kappa_\gamma|\leq 0.2$ can be 
achieved \cite{aih}. 
It is possible to place similar constraints on  $\cp$ violating and 
$\cp$ conserving anomalous couplings. 
Traditional studies of anomalous couplings \cite{tev,baur} have not discussed 
the coupling $\tilde{\kappa}_\gamma$ because there is a very strong indirect 
limit coming from the electric dipole moment of the neutron 
$|\tilde{\kappa}_\gamma|\leq 2\times 10^{-4}$ \cite{marciano}. This is 
a very tight indirect bound, and it indicates that it is quite 
unlikely that a non-zero $\cp$-odd effect will be observed in 
$p\bar{p}\ra W^\pm \gamma$. Nevertheless, in full generality, the 
electric dipole moment of the neutron and the $\cp$-odd observables that 
we study here, depend on different combinations of anomalous couplings and, 
therefore, complement each other. Because of this, and because 
the origin of $\cp$ violation is not understood, we would argue that 
an experimental search is necessary regardless of the merits of the limits that 
can be placed on couplings like $\tilde{\kappa}_\gamma$. 

Obviously, any 
experimental search will have to be able to distinguish between truly 
$\cp$ violating effects and possible $\cp$ biases of the detector. An 
important tool for this goal is the simultaneous measurement and comparison of 
as many observables as possible. With this in mind, we present in 
Figure~[1]  the observable $R_2(y_e)$ as an example. The curve corresponds 
to ${\tilde  \kappa}_\gamma=1$, and scales linearly with 
${\tilde  \kappa}_\gamma$. This observable is zero for all values 
of $y_e$ if $\cp$ is conserved.

\begin{figure}[htp]
\vspace*{-2.1cm}
\epsfxsize=12cm 
{\centerline{\epsfbox{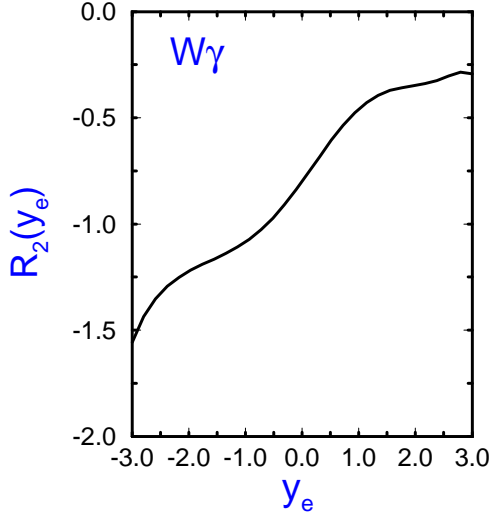}}}
\caption[]{The CP violating observable $R_2(y_e)/
{\tilde  \kappa}_\gamma$ for
$ p {\overline p} \rightarrow \ell^\pm \nu \gamma$ at
$\sqrt{S}=2~{\rm TeV}$.}
\label{figw}
\end{figure}

\section{$p\pp \ra Z \gamma \ra \ell^\pm \ell^\mp \gamma$}

Unlike the reaction which we studied in the previous section, 
$p\pp \ra Z \gamma \ra \ell^\pm \ell^\mp \gamma$ is self-conjugate 
under a $\cp$ transformation. This results in the need for different 
$\cp$-odd observables. Another difference between the two processes is that 
$p\pp \ra Z \gamma \ra \ell^\pm \ell^\mp \gamma$ does not receive contributions 
from Eq.~\ref{apefl} because 
the next to leading order effective Lagrangian for the symmetry breaking sector 
of the standard model does not produce anomalous $ZZ\gamma$ or $Z\gamma\gamma$ 
couplings. To generate a non-zero effect in this process we will have to go 
beyond the next to leading order effective Lagrangian, and we thus expect that 
any effect in $p\pp \ra Z \gamma \ra \ell^\pm \ell^\mp \gamma$ 
will be smaller than a corresponding effect in 
$p\pp \ra W^\pm \gamma \ra \ell \nu  \gamma$. 

In Ref.~\cite{hagi} the anomalous $Z(q_1^\alpha)
\gamma(q_2^\beta)V(P^\mu)$
vertices are parameterized in 
terms of several form factors, (where $V=Z,\gamma$).
 For example, there is a $\cp$ violating term 
of the form \cite{hagi}:
\begin{equation}
\Gamma^{\alpha\beta\mu}_{Z\gamma V}(q_1,q_2,P)={P^2-M_V^2 \over M_Z^2}
 h_1^V\bigl(q_2^\mu g^{\alpha\beta}-q_2^\alpha g^{\mu\beta}\bigr).
\label{hagiz}
\end{equation}
The overall factor $(P^2-M_V^2) / M_Z^2$ 
is required by electromagnetic gauge invariance
for $V=\gamma$ and by Bose symmetry for $V=Z$ \cite{hagi}. 
 Because of this factor, 
the contributions from Eq.~\ref{hagiz} to  $q\bar{q}\ra Z\gamma$  
are equivalent to those of  local operators of the forms:
\begin{eqnarray}
{\cal O}_\gamma &=& {1 \over \Lambda^2}{e g \over 2 c_\theta}
Q_q\bar{q}\gamma^\mu q Z^\nu F_{\mu\nu} \nonumber \\
{\cal O}_Z &=& {1 \over \Lambda^2}{e g \over 4 c_\theta}
\bar{q}\gamma^\mu [R_q(1+\gamma_5)+L_q(1-\gamma_5)] q Z^\nu F_{\mu\nu},
\label{hagizop}
\end{eqnarray}
where $F_{\mu\nu}$ is the electromagnetic field strength tensor, 
$L_q=\tau_3-2Q_q s_\theta^2$ and $R_q=-2Q_q s_\theta^2$.
We have changed the normalization scale from $M_Z$ to $\Lambda$, the 
scale of the new physics responsible for this operator. 
Since we want to insist on an effective Lagrangian that preserves the 
symmetries of the standard model we must convert 
these operators into fully $SU(2)_L\times U(1)_Y$
gauge invariant versions.
 There are at least two ways to do this. One way 
would be to obtain the $Z^\nu$ field from an $SU(2)$ covariant derivative 
acting on the fermion field. There are several dimension~6 operators with 
the desired properties listed in Ref.~\cite{buchmuller}. 
One of them is:
\begin{eqnarray}
{\cal L} &=& i {\tilde{\alpha} \over 2 \Lambda^2}\bigl(
{\cal O}_{qB}+{\cal O}_{uB}+{\cal O}_{dB}\bigr)+{\rm ~h.c.}\nonumber \\
&=& i{\tilde{\alpha} \over \Lambda^2}\bar{q}\gamma_\mu 
\biggl(\partial_\nu+ieQ_qA_\nu 
+ {ig\over 4c_\theta}[L_q(1-\gamma_5)+R_q(1+\gamma_5)]Z_\nu\biggr)qB^{\mu\nu} 
+{\rm ~h.c.}
\label{invbu}
\end{eqnarray}
With this approach we have an operator that is suppressed by only 
two powers of the new physics scale. However, as one can see from 
Eq.\ref{invbu}, the fully gauge invariant version of the operator 
Eq.~\ref{hagizop} also generates $\bar{q} qZ$ and $\bar{q}q\gamma$ vertices. 
When all of these new vertices are systematically taken into account, 
the interference between the lowest order standard model amplitude and 
the new $\cp$-violating amplitude for the process 
$\bar{q}q\ra \ell^+\ell^-\gamma$ 
vanishes and one is left with no $\cp$-odd effects. This is a rather 
surprising result that we have checked for all the relevant operators 
in Ref.~\cite{buchmuller}.\footnote{
Similar observations have been made before in the literature \cite{rujula}.}

A second way to make Eq.~\ref{hagizop} fully gauge invariant 
uses a non-linearly realized electroweak symmetry breaking 
sector. In this case we think of Eq.~\ref{hagizop} as the unitary gauge 
version of fully gauge invariant operators that can be constructed as 
described in Refs.~\cite{peccei,chano}. This construction 
generates the $\cp$ odd observables which we discuss next. However, 
the counting rules appropriate for this construction 
\cite{georgi}, 
tell us that the operators are suppressed by {\it four} inverse powers of the 
symmetry breaking scale and any effects are, therefore, expected to be 
extremely small. 

Taking the unitary gauge Lagrangian 
\begin{equation}
{\cal L} = {v^2 \over \Lambda^4}\biggl( h_1^\gamma {\cal O}_\gamma 
+h_1^Z {\cal O}_Z\biggr)
\end{equation}
we find for $\bar{q} q \rightarrow \ell^+ \ell^- \gamma$ 
a term linear in $h_1^V$ that contains a triple product:
\begin{eqnarray}
|{\cal M}_{CP}|^2 &=&
{1\over 12} {v^2 h_1^V\over \Lambda^4}{g^4 \over  c^4_\theta}
e^2 Q_q \biggl({1\over m^2_{\ell \ell} -M^2_Z}\biggr)^2
\epsilon(p_q,p_{\bar{q}},p^+,p^-)  \nonumber \\ &\cdot &
\biggl[(L_\ell^2+R_\ell^2)C_{V1}\biggl(
{p_q\cdot(p^--p^+)\over 2p_q \cdot p_\gamma}-
{p_{\bar{q}}\cdot(p^--p^+)\over 2p_{\bar{q}} \cdot p_\gamma}\biggr)
\nonumber \\ &+&
(L_\ell^2-R_\ell^2)C_{V2}\biggl(1-
{p_q\cdot p_{\bar{q}}\over 2p_q \cdot p_\gamma}-
{p_q\cdot p_{\bar{q}}\over 2p_{\bar{q}} \cdot p_\gamma}\biggr)\biggr]
\end{eqnarray}
where $C_{\gamma 1}=Q_q(L_q-R_q)$, $C_{Z 1}=L_q^2-R_q^2$,
$C_{\gamma 2}=Q_q(L_q+R_q)$ and $C_{Z 2}=L_q^2+R_q^2$.
We use the narrow-width approximation  again 
and neglect the contributions from radiative $Z$ decays. As 
before, we restrict ourselves to the region of phase space where this 
approximation works best by imposing the cuts: 
$p_{T\gamma} > 10~{\rm GeV}$, $p_{T\ell} > 20~{\rm GeV}$,  
$|y_\gamma| < 2.4$, $|y_\ell|<3.0$, $\Delta R(\ell\gamma) > 0.7$ and 
$M_{\ell^+\ell^-\gamma} > 100~{\rm GeV}$. As in the case of $W^\pm \gamma$, 
the first and last two cuts, defined as in Ref.~\cite{baur}, suppress the 
radiative $Z$ decays and the remainder
of the cuts are typical Tevatron 
acceptance cuts \cite{tev}. 

In this reaction there can also be an absorptive phase in the form-factor 
of Eq.~\ref{hagiz}. If we include this absorptive phase, 
$\sin\delta(qqZ\gamma)$, we find additional $\cp$ violating contributions to 
the differential cross-section. 
As in the case of $W^\pm \gamma$, these new 
contributions are too cumbersome to write out explicitly but we include them 
in our numerical work. 

The reaction $p\bar{p}\ra \ell^+ \ell^- \gamma$ is self-conjugate under a $\cp$ 
transformation and this allows us to write fully integrated $\cp$ 
odd observables in the lab-frame such as:
\begin{eqnarray}
A_T & \equiv & \int {\rm sign}(\vec{p}_{beam}\cdot
(\vec{p}_{\ell^+}\times \vec{p}_{\ell^-})) d\sigma \nonumber \\
A_E & \equiv & \int  {\rm sign}(E_{\ell^+}-E_{\ell^-})d\sigma 
\label{integob}
\end{eqnarray}

\begin{figure}[htp]
\vspace*{-2.1cm}
\epsfxsize=12cm
{\centerline{{\epsfbox{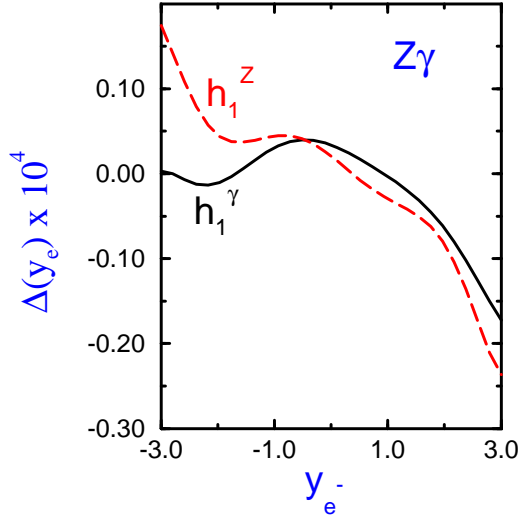}}}}  
\vspace*{-.3cm}  
\caption[]{$\Delta(y_\ell)$ for $p {\overline p}\rightarrow
\ell^+\ell^- \gamma$ at $\sqrt{S}=2~TeV$.  These curves are normalized
to $h_1^V=1$ and scale linearly with $h_1^V$.  We have used 
$\Lambda=1~TeV$. }
\label{figz}
\end{figure}

There will also be observables analogous to those in 
Eqs.~\ref{cpobst},~\ref{cpobs}. As an illustration we present in 
Figure~\ref{figz} the following observable:
\begin{equation}
\Delta(y_0)  \equiv 
{ {d {\sigma}\over dy_{\ell^+} } \mid_{y_{\ell^+}=y_0}-
{d {\sigma}\over dy_{\ell^-} }\mid_{y_{\ell^-}= -y_0} \over
{d {\sigma}\over dy_{\ell^+} }\mid_{y_{\ell^+}=y_0} +
{d {\sigma}\over dy_{\ell^-} }\mid_{y_{\ell^-}= -y_0} } 
\label{cpobsz}.
\end{equation}
Numerically we use $\sin\delta(qqZ\gamma)=1$ and $\Lambda=1$~TeV. 

With the set of cuts that we are using we find numerically that:
\begin{eqnarray}
\sigma &\approx & 348 ~fb
\nonumber \\
A_T &\approx & 
(5.6 h_1^\gamma + 12 h_1^Z)\times 10^{-3}
\biggl({1~{\rm TeV}\over \Lambda}\biggr)^4 ~fb
\nonumber \\
A_E &\approx & 
(1.3 h_1^\gamma + 0.54 h_1^Z)\times 10^{-3}
\biggl({1~{\rm TeV}\over \Lambda}\biggr)^4 \sin\delta(qqZ\gamma)~fb
\label{zresult}
\end{eqnarray}
For an integrated luminosity of $10fb^{-1}$ we thus expect about 
3500 $Z \gamma$ events within the phase space region defined by our cuts. 
This translates into $95\%$ confidence level bounds:
\begin{equation}
|h_1^\gamma + 2 h_1^Z| \leq 2700 
\biggl({\Lambda \over 1~{\rm TeV}}\biggr)^4.
\label{zbound}
\end{equation}

\section{Conclusions}

We have constructed several $\cp$-odd asymmetries that can be used
in the
search for $\cp$ violation in $W^\pm \gamma$ and $Z\gamma$ events in
$p\pp$ colliders. We have estimated the contributions to these
asymmetries from some simple $\cp$ violating effective operators that
respect the symmetries of the Standard Model. 
We find that an upgraded Tevatron with $10~fb^{-1}$ at $\sqrt{S}=2~{\rm TeV}$
can place limits of $\mid \alpha_{13}+\alpha_{14}\mid < 4 
(\Lambda/1~{\rm TeV})^2$ 
and $\mid h_1^\gamma+2 h_1^Z\mid < 2700 (\Lambda/1~{\rm TeV})^4$. 
The first bound corresponds to 
$\mid {\tilde \kappa}_\gamma\mid < .1$, which is comparable to
the bound obtainable for the $\cp$ conserving couplings in the next-to-leading 
effective Lagrangian.

\vspace{1in}

\noindent {\bf Acknowledgements} The work of G.V. was supported in
part by the DOE OJI program under contract number DEFG0292ER40730. G. V. thanks
the theory group at the University of Melbourne
and X-G.H thanks the theory group at BNL  
for their hospitality while part of this work
was performed. The work of X-G.H. was supported by the Australian Research 
Council. We are grateful to U. Baur, S. Errede, and J.P. Ma for
useful conversations.

\end{document}